\documentclass[prl,amssymb,superscriptaddress,twocolumn]{revtex4-2}
\usepackage{amsmath}
\usepackage{amssymb}
\usepackage{graphicx}
\usepackage{ulem}
\usepackage{mathtools}
\usepackage{mathrsfs}
\usepackage[colorlinks=true,linkcolor=blue,citecolor=blue,filecolor=blue,urlcolor=blue,pdfstartview=FitH,breaklinks=true]{hyperref}
\usepackage{xcolor}
\usepackage{ulem}

\newcommand{\be}{\begin{equation}}
\newcommand{\ee}{\end{equation}}

\newcommand{\lf}{\left}
\newcommand{\rg}{\right}
\newcommand{\ra}{\rangle}
\newcommand{\la}{\langle}

\newcommand{\bea}{\begin{eqnarray}}
\newcommand{\eea}{\end{eqnarray}}

\usepackage{xcolor}
\usepackage{color}

\begin{document}

%\title{Non-Abelian Thouless pumping in a $SO(5)$ ladder} \\
\title{Non-Abelian Thouless pumping in a Rice-Mele ladder}

\author{Carlo Danieli}
\email{carlo.danieli@cnr.it}
\affiliation{Istituto dei Sistemi Complessi, Consiglio Nazionale delle Ricerche, Via dei Taurini 19, 00185 Rome, Italy}
\author{Valentina Brosco}
\email{valentina.brosco@cnr.it}
\affiliation{Istituto dei Sistemi Complessi, Consiglio Nazionale delle Ricerche, Via dei Taurini 19, 00185 Rome, Italy}
\affiliation{Dipartimento di Fisica, Universit\`a ``La Sapienza'', P.le A. Moro, 5 (00185) Roma, Italy}
\author{Laura Pilozzi}
\affiliation{Istituto dei Sistemi Complessi, Consiglio Nazionale delle Ricerche, Via dei Taurini 19, 00185 Rome, Italy}
\affiliation{Centro Ricerche Enrico Fermi, Piazza del Viminale, 1, I-00184 Rome, Italy}
\author{Roberta Citro}
\affiliation{Dipartimento di Fisica “E.R. Caianiello”, Universit\`a di Salerno, Via Giovanni Paolo II, 132, I-84084 Fisciano (SA), Italy}

\date{\today}

\begin{abstract}

Non-Abelian Thouless pumping intertwines adiabatic quantum control and topological quantum transport and it holds potential for quantum metrology and computing. In this work, we introduce a ladder model featuring two doubly-degenerate bands and we show that adiabatic manipulation of the lattice parameters results in non-Abelian Thouless pumping, inducing both the displacement of an initially localized state and a geometric unitary transformation within the degenerate subspace. Additionally,  we show that the structure and symmetry of the ladder model can be understood through its connection to a Yang monopole model.  The proposed Hamiltonian can be realized using cold atoms in optical lattices, enabling the experimental demonstration of non-Abelian Thouless pumping in a genuinely quantum many-body system.

\end{abstract}

\maketitle

\begin{figure}[t!]
    \centering
    \includegraphics[width=\columnwidth]{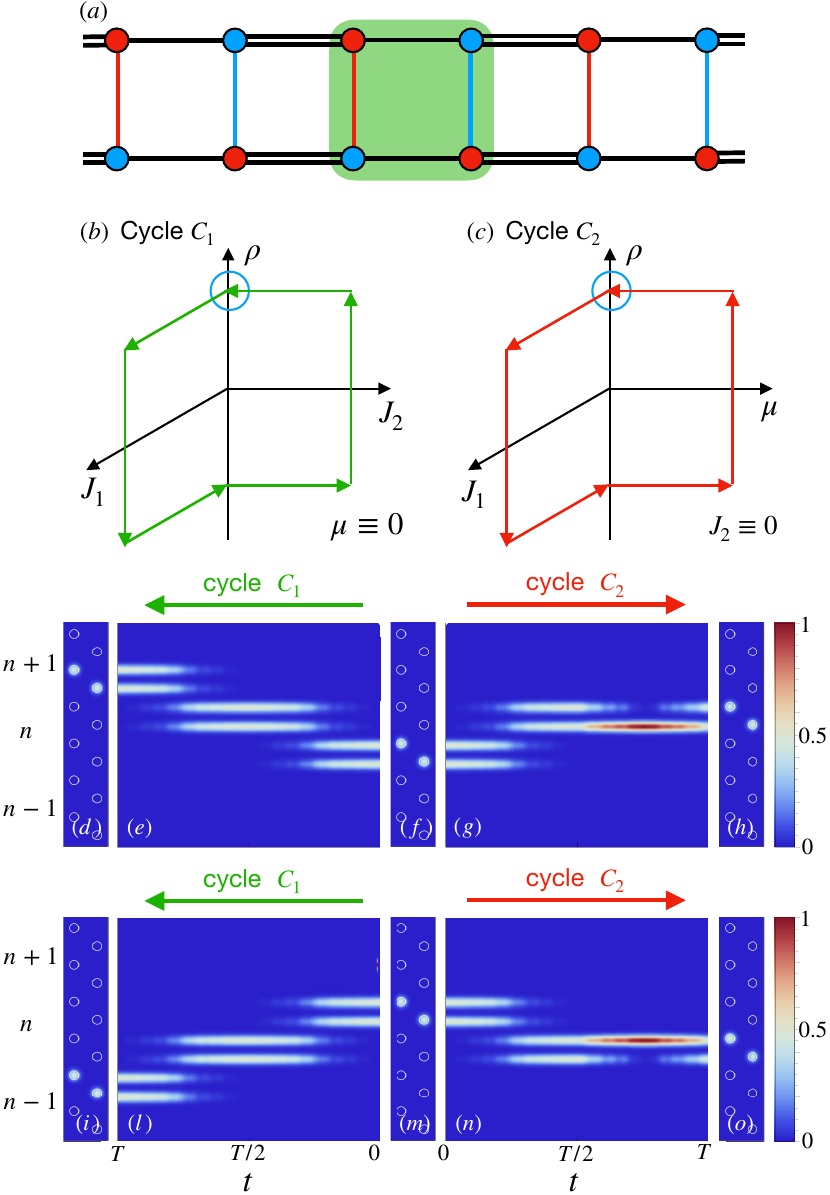}
    \caption{(a) Ladder in Eq.~\eqref{eq:H} where the green box indicates a unit-cell. Red and blue sites have opposite on-site potential $\mu$. Single and double lines indicate the tunnelings $J_1$ and $J_2$ while red and blue vertical lines denote the inter-chain tunnelling $\rho$ and $-\rho$. 
    (b-c) Pumping cycles $C_1$ and $C_2$ respectively in the parameter space. %$\{\rho,\mu, J_1,J_2\}$. 
    The blue circles indicate the initial point.
    (d)–(h) Numerically evaluated field’s intensity along $C_1$ and $C_2$ with initial state $v_{a,n_0}^+$. (f) shows the initial excited $v_{a,n_0}^+$. (e) and (g) show the evolution according to $C_1$ and $C_2$ respectively. 
    (d) and (h) show the field intensity in the final states after one pumping cycles $C_1$ and $C_2$. 
    (i)-(o) Same as (d)–(h) with initial state $v_{b,n_0}^+$ 
    }
    \label{fig:model}
\end{figure}

An important geometric aspect of quantum mechanics emerges when the Hamiltonian of a quantum system varies adiabatically and cyclically with time. In this case, the evolution operator over a cycle yields a transformation that depends solely on the topological structure of the Hilbert space and the geometry of the cycle while being independent of dynamical details such as the energy levels or the cycle duration.
When the adiabatic evolution involves a non-degenerate eigenstate the geometric part of the evolution coincides with the Berry's phase~\cite{Berry1984}, conversely, for a $N$-degenerate eigenstate the geometric evolution is a $U(N)$ transformation, called non-Abelian holonomy~\cite{Wilczek1984}. 
Beyond their significance in adiabatic quantum evolution, geometric phases and holonomies are crucial for understanding the properties of Bloch bands in solids~\cite{wilczek1989}. They underlie polarization theory and many fascinating phenomena, such as quantum Hall effect~\cite{thouless1982}, the spin Hall effect~\cite{murakami2004}, and topological phases~\cite{qi2011}.

The interplay of geometry, lattice symmetries, and adiabatic dynamics emerges particularly in Thouless pumping~\cite{thouless1983}. Thouless pumping refers to transport induced by the adiabatic and cyclic manipulation of a lattice potential in the absence of any external bias. Under suitable conditions, this phenomenon yields topologically quantized transport, enabling the direct measurement of topological invariants~\cite{thouless1983}. Thouless pumping has been experimentally realized in various systems~\cite{citro2023}, including cold atoms  and spin in optical lattices~\cite{lohse2015, nakajima2016,schweizer2016}, and photonic waveguide arrays~\cite{kraus2012}. It can be employed to explore the breakdown of topological phenomena in the presence of interactions~\cite{stenzel2020, ke2017, nakagawa2018, liu2023, walter2023, viebahn2024}, disorder~\cite{cerjan2020,huang2024}, nonlinearities~\cite{mostaan2022, jung2022}, or dissipation~\cite{ravets2024}, leading to fractional topological quantization~\cite{mostaan2022, juergensen2023, marra2015,marra2017,taddia2017} and topological phase transitions. Recent theoretical work~\cite{brosco2021} has demonstrated that Thouless pumping can exhibit non-Abelian characteristics in systems with degenerate Bloch bands. Subsequently, non-Abelian Thouless pumping has been implemented in photonic~\cite{sun2022, pilozzi2022} and acoustic waveguide arrays~\cite{you2022}. In these setups, the propagation of electromagnetic waves is effectively described by an Hamiltonian having a tripod structure~\cite{brosco2021, brosco2021a} and featuring a doubly-degenerate flat band. Tripod Hamiltonians have long been studied in relation to non-Abelian holonomies in atomic transitions~\cite{duan2001}, superconducting nanocircuits~\cite{faoro2003}, Cooper pair pumps~\cite{brosco2008}, and more recently photonic systems~\cite{neef2022}.

In the present work, we envisage a non-Abelian Thouless pump in a lattice with two dispersive, doubly degenerate bands -- hence moving beyond the paradigm of tripod flat-band systems discussed in~\cite{brosco2021}. 
The Hamiltonian has a ladder structure and can be implemented by using cold atoms in optical lattices, thereby enabling the demonstration of non-Abelian Thouless pumping in an inherently quantum many-body system.
Furthermore, as we show below, its structure and symmetry properties~\cite{demler1999} can be explained through a relation with a Yang monopole model~\cite{yang1954,yang1978}. 
Thanks to their exceptionally high level of control and robustness, non-Abelian Thouless pumps hold significant promise for applications in quantum computing~\cite{zanardi1999,zhang2023geometric}, routing~\cite{bottarelli2023} and  metrology~\cite{klitzing2017,jeckelmann2018,yu2024}. Our work therefore has both a practical and fundamental relevance as it paves to the development of different holonomic devices and the investigation of the interplay between the geometric and dynamical properties of many-body quantum systems.

The Hamiltonian describes two coupled Rice-Mele~\cite{rice1982} chains and it can be cast as follows:
\be 
\small
\begin{split}
\hspace{-10mm}
H &= \sum_{n}\sum_{M=U,D} \lf[J_1\, a_{n, M}^\dagger b_{n, M}+J_2\, a_{n,M}^\dagger b_{n-1,M}+ {\rm H.c.}\rg]\\
&+ \mu  \sum_{n}  \lf[a_{n, U}^\dagger a_{n,U} - a_{n, D}^\dagger a_{n,D}   - b_{n, U}^\dagger b_{n,U} + b_{n, D}^\dagger b_{n,D}  \rg] \\
&+\rho  \sum_{n}  \lf[ a_{n, U}^\dagger a_{n,D}- b_{n, U}^\dagger b_{n,D} +{\rm H.c.}\rg] 
\end{split}
\label{eq:H}
\ee
where $ c_{n,M}^\dagger$ and $ c_{n,M}$ (with $c=a,b$) are the creation and annihilation operators of sublattice $a,b$ in the unit cell $n$ and chain $M=U,D$. 
In  Eq.~\eqref{eq:H} $J_1$ and $J_2$ represent the intra- and inter-cell hopping along the two chains, $\mu$ is a staggered on-site potential and $\rho$ is a staggered interchain coupling.
%The third line yields a staggered interchain hopping with amplitude  $\rho$. 
The model is schematically illustrated in Fig.~\ref{fig:model}(a).%, where the four sites unit-cell $u_n$ is colored in green. 

%We recast  write the Hamiltonian $H$ in Eq.~\eqref{eq:H} in $k$-space. For shortness, we group the creation and annihilation operators in a
Introducing the  four-dimensional spinor  creation and annihiliation operator %spinors 
$\Psi_k^{(\dagger)}=(a_{k,U}^{(\dagger)},a_{k,D}^{(\dagger)},b_{k,U}^{(\dagger)},b_{k,D}^{(\dagger)})$, we can recast the 
%and use the Pauli matrices $\sigma_j$ and $\tau_j$ and the identity matrices $\sigma_0$ and $\tau_0$.
 Hamiltonian $H$ in momentum space as 
\be 
H= \sum_k \Psi_k^\dag \lf[  \lf(J_x \tau_x +J_y \tau_y\rg) \otimes\sigma_0 
+ \tau_z\otimes \lf(\rho \, \sigma_x+\mu\,\sigma_z \rg)\rg]\Psi_k.
\label{eq:H_k}
\ee
Here, $J_k = J_1 + J_2 e^{ik}$ is decomposed as the sum of  $J_x=J_1+J_2\cos k$ and  $J_y=J_2 \sin k$ and  $\sigma_j$ and $\tau_j$ are Pauli matrices spanning the spin and site indices.
The Bloch spectrum of the ladder consists of two doubly-degenerate bands with dispersion, 
\be E_{\pm}(k)=\pm \sqrt{\mu^2 + \rho^2 + |J_k|^2}\equiv \pm \Delta\ee corresponding to the Bloch states~\cite{Supple}
\bea
%\small
%\begin{split}
\hspace{-2.5mm} &&|\psi_{a}^{\pm}(k)\ra = \frac{1}{\mathcal{R}_\pm} \lf[\rho |a_{k,U}\ra + (-\mu \pm \Delta ) |a_{k,D}\ra+J_k
 |b_{k,D}\ra\rg] 
 \label{eq:psi0}\\ 
\hspace{-2.5mm} &&|\psi_{b}^{\pm}(k)\ra =  \frac{1}{\mathcal{R}_\pm} \lf[ - J_k^* |a_{k,U}\ra + (\mu \mp  \Delta ) |b_{k,U}\ra + \rho |b_{k,D}\ra\rg]  
%\end{split}
\label{eq:psi1}
\eea
so that $H |\psi_{m}^{\pm}(k)\ra=E_{\pm}(k)|\psi_{m}^{\pm}(k)\ra$
with $m=a,b$  and  
$\mathcal{R}_\pm = 1/\sqrt{2 \Delta ( \Delta \mp  \mu)}$. 
The spectrum is thus gapless for $\mu=\rho=0$ and $J_1=J_2$ and gapped otherwise. \\

%{\it Non abelian Thouless pumping} -- 
Thouless pumping is achieved by modulating periodically and adiabatically at least two parameters  defining the Hamiltonian $H$.
The non-Abelian nature of the evolution implies that pumping cycles cannot only shift but also geometrically manipulate bond and plaquette states along the ladder. 
At time $t = 0$, we initialize the system in a Wannier state defined with coefficients $c_{\nu}$, 
$|\psi_{n_0}^{\pm}(0)\rangle = 
\sum_{k,\nu} c_{\nu} |\psi_{\nu}^{\pm}(k)\ra e^{i k n_0}$ 
belonging to one of the two bands  
$E_\pm$, and localized within the unit cell $n_0$. 
Following Ref.~\cite{brosco2021}, in the adiabatic regime the evolution of $|\psi_{n_0}^{\pm}(0)\rangle$ can be expressed as follows 
\be
%\small
|\psi_{n_0}^{\pm}(T)\rangle 
= 
\sum_{k\nu\eta}
 c_{\nu }\left[ W^\pm(0,T)\right]_{\eta\nu}|\psi_{\eta}^{\pm}(k)\ra  e^{i k n_0} 
\label{eq:propagation1}
\ee
where $T$ denotes the driving period and the adiabatic evolution operator is given  by~\cite{brosco2021} 
\be \label{eq:adev} W_\pm(0,T) = e^{ i\theta_{\rm d}^{\pm}} \mathcal{P} \exp\left[ i\int_{0}^{T} \Gamma^\pm_{t} dt \right].\ee 
In the above equation, $\theta_{\rm d}^{\pm}$ denotes the dynamical phase $\theta_{\rm d}^{\pm} = \int_{0}^{T} E_\pm (t) dt$  
while the geometric part of $W_\pm$  is given by  a path ordered exponential $\mathcal{P}$  of the Wilczek-Zee connection
$\left[\Gamma^\pm_{t}\rg]_{\nu\nu'} = \la \psi_{\nu}^{\pm}(k)  | \partial_t |\psi_{\nu'}^{\pm}(k)\ra $ associated to the two bands~\cite{Wilczek1984}.

%The connection $\Gamma^\pm_{t}$ reduces to the Berry connection for nondegenerate systems
%, and it demonstrates that by adiabatically manipulating the lattice's parameters, single-particle gates can be achieved through geometric transformations. 
%
%This path integral is then multiplied by a 
%
%For the Hamiltonian \( H \) in Eq.~\eqref{eq:H}, the Wilczek-Zee connection associated with time-dependent variations in the different parameters can be expressed as

Starting from Eqs.(\ref{eq:psi0},\ref{eq:psi1}) we can express the connection $\Gamma^\pm_{t}$ generated by time-dependent drivings on the different parameters of the Hamiltonian \( H \) as follows
%The Wilczek-Zee connection associated with the Hamiltonian \( H \) in Eq.~\eqref{eq:H}, when time-dependent variations are introduced in the parameters, can be expressed as
%
%For the Hamiltonian $H$ in Eq.~\eqref{eq:H} the Wilczek-Zee connection generated by enforcing time-dependent drivings on the different parameters can be expressed as~\cite{Supple}
\be
\begin{split}
\hspace{-2.5mm}
\Gamma^\pm_{t}&=
\frac{1}{\mathcal{R}_\pm^2}
\lf[\lf(J_2 \Dot{J_1}-J_1\Dot{J_2}\rg) \sin k\,\hat\sigma_z  
+(\Dot{J_1}\rho-J_1\Dot{\rho}) \hat\sigma_y \rg.\\
&\qquad\quad \lf. + (\Dot{J_2}\rho-J_2\Dot{\rho})( \cos k\, \hat \sigma_y - \sin k\, \hat \sigma_x)  \rg].
\end{split}
\label{eq:holonomy_t}
\ee
where $\hat \sigma_j$ are the Pauli matrices in the basis of the degenerate eigenstates -- see~\cite{Supple} for details.
Following Ref.~\cite{brosco2021} we can express the displacement of the state $|\psi_{n_0}^{\alpha}(0)\rangle$ as 
\be \label{eq:displ}
 \Delta x=\sum_{\nu\mu}c^*_\nu c_\mu D^{\alpha}_{\mu\nu},%\frac{1}{2\pi} \int_{-\pi}^{ \pi}\partial_k \theta_{\rm d}^{\pm}   dk +
\ee
 where the displacement matrix $D^{\alpha}_{ab}$ can be recast as
\be\label{eq:displ-matr}D^{\alpha}_{\mu\nu}=\frac{1}{2\pi} \int_{0}^{T}\!\!\!dt \int_{-\pi}^{ \pi}  \!\! \!\!  dk \lf[
W^\dag_\alpha\,{\cal F}^\alpha_{kt}W_\alpha \rg]_{\mu\nu} \ee
with $\alpha=\pm$, ${\cal F}^\alpha_{kt}=\partial_k \Gamma_t^\alpha-\partial_t \Gamma_k^\alpha+i\lf[\Gamma_t^\alpha, \Gamma_k^\alpha\rg]$ denoting the non-Abelian field strength  matrix %in vector notation, i.e. $|\Psi_{\nu}\ra=(|\psi_{\nu1}\ra,..., |\psi_{\nu d_\nu}\ra )$ where $d_\nu$ is the dimension of the degenerate subspace, 
and $\Gamma^\pm_{k}$ indicating the $k$-connection $\left[\Gamma^\pm_{k}\rg]_{\nu\nu'} = \la \psi_{\nu}^{\pm}(k)  | \partial_k |\psi_{\nu'}^{\pm}(k)\ra $.
Equations (\ref{eq:displ},\ref{eq:displ-matr}) illustrate the topological and geometrical significance of non-Abelian Thouless pumping. %, particularly for pumping cycles involving multiple degenerate Bloch bands. 
In these regards, a particularly intriguing aspect of this phenomenon, rooted in its geometric nature, is the exceptional level of control it offers over both the state's evolution and the transport process. By suitably designing the pumping cycles we can indeed engineer different combination of translations along the lattice and rotation in the degenerate subspace.

In the construction of the pumping cycles, we impose two conditions: 
(i) $\min_{k,t} |E_{+}(k) - E_{-}(k)| T \gg 1$, and 
(ii) $\max_{k,t}\lf| \partial_k E_{\pm}(k) \rg|  T \ll a$, with $a$ denoting the lattice spacing.  
Condition (i) relates the driving period $T$ with the band-gap, and it expresses the adiabaticity criterion. 
Instead, condition (ii) relates $T$ with the group velocity of a band, and it requires that the displacement generated by the pumping in one cycle, typically of the order of one unit cell, is much smaller than the dynamically induced dispersion~\cite{Supple,hu2019,liu2024}.
In flat band systems~\cite{brosco2021} condition (ii) is always fulfilled as $\partial_k E_{\pm}(k) =0$.  
Conversely, when the bands are not flat, 
%These conditions (i) and (ii) are enforced for every $k$ in the Brillouin zone and for every time $t$, as modulating the ladder parameters deforms the bands $E_\pm$. 
satisfying simultaneously both inequalities guarantees that the pumping is {\it adiabatic} and {\it weakly dispersive}.
As detailed in~\cite{Supple}, there exist a wide region in parameters regions where  both conditions (i) and (ii) are satisfied.

%{\it Pumping cycles} -- 
We consider two pumping cycles, called $C_1$ and $C_2$. 
As common starting point of these cycles at $t=0$, we choose $J_1=J_2=\mu=0$ and $\rho=\rho_0\neq0$. 
This choice reduces the ladder in Fig.~\ref{fig:model}(a) to a set of decoupled dimers, as only the transversal hopping is present. 
As prescribed by Eqs.~(\ref{eq:psi0},\ref{eq:psi1}), we initialize the system in a Wannier state $|\psi_{n_0}^{\pm}(0)\rangle$ localized in the  unit-cell $n_0$ belonging to the $\pm$ bands. We set:
$|v_{a,n_0}^\pm\ra  = \frac{\delta_{n,n_0}}{\sqrt{2}} \lf[ |a_{n,U} \ra \pm |a_{n,D}\ra  \right]$ and 
$|v_{b,n_0}^\pm\ra  = \frac{\delta_{n,n_0}}{\sqrt{2}} \lf[ |b_{n,U} \ra \mp |b_{n,D}\ra  \right]$.

%\bea
%\label{eq:IC1a}
%v_{a,n_0}^\pm  &=& \frac{\delta_{n,n_0}}{\sqrt{2}} \lf[ |a_{n,D} \ra \pm |a_{n,U}\ra  \right]  \\
%\label{eq:IC1b}
%v_{b,n_0}^\pm  &=& \frac{\delta_{n,n_0}}{\sqrt{2}} \lf[ |b_{n,D} \ra \mp |b_{n,U}\ra  \right].
%\eea
%
The cycles $C_1$ and $C_2$ are schematically depicted in Fig.~1(b,c). During both cycles, $J_1$ and $\rho$ change adiabatically. However, in cycle $C_1$, the onsite potential remains zero, while $J_2$ varies adiabatically. Conversely, in cycle $C_2$, $J_2$ is set to zero, and $\mu$ undergoes a variation.
%Cycle $C_2$ involves the same linear manipulations as $C_1$. In this case however, the onsite energy $\mu$ is pumped instead of the longitudinal hopping $J_2(t)\equiv0$. 
The Wilson loops $W^\pm$ entering the adiabatic evolution operator defined in  Eqs.~(\ref{eq:adev},\ref{eq:holonomy_t}) can be calculated analytically using the Wilczek-Zee connection %defined in Eq. \eqref{eq:holonomy_t}  
and, up to phase factors, lead to~\cite{Supple}
\be
W^\pm_{C_1} =%(0,T) =
 \begin{pmatrix}
 e^{ik} & 0\\
0  & e^{-ik}
 \end{pmatrix} 
  \qquad
W^\pm_{C_2} =%(0,T) = 
 \begin{pmatrix}
 0 & 1\\
-1  & 0 
 \end{pmatrix} 
\label{eq:W_C1_C2}.
\ee
Via Eq.~\eqref{eq:propagation1} we obtain that, over one period $T$ cycle, $C_1$ shifts the states of one unit-cell to their right/left respectively -- {\it i.e.} the excitations 
$v_{a,n_0}^\pm$ and $v_{b,n_0}^\pm$ are shifted to $v_{a,n_0+1}^\pm$ and $ v_{b,n_0-1}^\pm$ respectively~\cite{Supple}. 
Cycle $C_2$ on the other hand swaps these state within one unit-cell  -- {\it i.e.} the excitations 
$v_{a,n_0}^\pm$ and $v_{b,n_0}^\pm$ are mapped to 
$v_{b,n_0}^\pm$ and $ -  v_{a,n_0}^\pm$ respectively~\cite{Supple}. 
In other words, cycles $C_1$ and $C_2$ generate chiral quantized change displacement along the ladder, as shown in Fig. 2.

\begin{figure}[t!]
    \centering
    \includegraphics[width=\columnwidth]{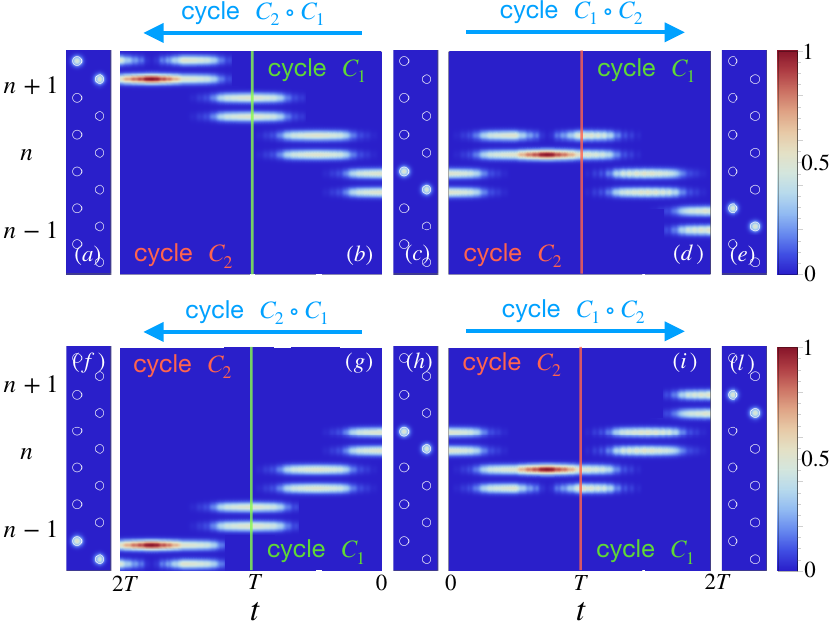}
    \caption{(a)-(e) Numerically evaluated field’s intensity along $C_2\circ C_1$ and $C_1\circ C_2$ starting from state $v_{a,n_0}^+$. (c) shows the initial excited $v_{a,n_0}^+$. (b) and (d) show the evolution according to $C_2\circ C_1$ and $C_1\circ C_2$ respectively. (a) and (e) show the field intensity in the final states after one pumping cycles $C_2\circ C_1$ and $C_1\circ C_2$. 
    (f)-(l) Same as (d)–(h) with initial condition $v_{b,n_0}^+$. 
    }
    \label{fig:nonabelian}
\end{figure}
The numerical results in Fig.~\ref{fig:model}(d-o) obtained by solving the Schr\"odinger equations 
$i\partial_t|\psi\ra=H(t)|\psi\ra$ of the ladder are in agreement with the analytical prediction obtained in  Eqs.~(\ref{eq:propagation1},\ref{eq:holonomy_t},\ref{eq:W_C1_C2}).  
Due to symmetry $E_-=-E_+$ of the Bloch bands, we focus on the positive band $E_+$ and show results only for the states $v_{a,n_0}^+$ and $v_{b,n_0}^+$. 
Along both $C_1$ and $C_2$, the conditions (i) and (ii) for adiabatic and weakly dispersive pumping reduce to $\frac{1}{2 \rho_0}  \ll T <  +\infty$~\cite{Supple}. In our numerical tests we set $\rho_0 = 0.05$, which results in $T \gg 10$ -- hence, we choose $T = 10^3$. 
Fig.~\ref{fig:model}(f) and (m) respectively show the initially states $v_{a,n_0}^+$ and $v_{b,n_0}^+$. 
Their propagation along one period of cycle $C_1$ are shown in Fig.~\ref{fig:model}(e) and (l), and their final state in Fig.~\ref{fig:model}(d) and (i). 
Likewise, the propagation of $v_{a,n_0}^+$ and $v_{b,n_0}^+$ along one period of cycle $C_2$ are shown in Fig.~\ref{fig:model}(g) and (n), and the correspondent final state in Fig.~\ref{fig:model}(h) and (o).
Focusing at first on the time evolution along cycle $C_1$ -- {\it i.e.} Fig.~\ref{fig:model}(e,l) --  we notice that in the first half of the cycle, namely $0\leq t\leq \frac{T}{2}$, where only the hoppings $J_1$ and $\rho$ are activated, $v_{a,n_0}^+$ and $v_{b,n_0}^+$ are shifted within the unit-cell. 
Then, in the second half of the cycle, namely $ \frac{T}{2} \leq t\leq T$,  where only the hoppings $J_1$ and $J_2$ are activated, $v_{a,n_0}^+$ and $v_{b,n_0}^+$ are shifted to the neighboring unit-cell.
This holds analogously for cycle $C_2$-- {\it i.e.} Fig.~\ref{fig:model}(g,n). Indeed, in the first half of the cycle,  where only $J_1$ and $\rho$ are activated, $v_{a,n_0}^+$ and $v_{b,n_0}^+$ are shifted within the unit-cell.  
Then in the second half of the cycle, where only the hopping $\rho$ and the potential $\mu$ are activated, the states $v_{a,n_0}^+$ and $v_{b,n_0}^+$ are rotated.

These two cycles therefore yield different chiral quantized displacement, as in $C_1$ we set $\mu(t)\equiv0$ to avoid state rotation, while in $C_2$  we set $J_2(t)\equiv0$ to prevent transport along the chains. 
Furthermore, they allow to unravel the non-Abelian nature of the Thouless pumping
%in this system Eq.~\eqref{eq:H}. 
%Indeed, 
since their holonomies do not commute, {\sl i.e.} $C_2\circ C_1\neq C_1\circ C_2$.
This is shown in Fig.~\ref{fig:nonabelian} for the inital states $v_{a,n_0}^+$ and $v_{b,n_0}^+$.
%, respectively plot in panels (c) and (h). 
Following $C_2\circ C_1$  the states $v_{a,n_0}^+$ and $v_{b,n_0}^+$ are first shifted to their neighboring unit-cells and then swapped. %within the unit-cell by cycle $C_2$.
%The final states are shown in Fig.~\ref{fig:nonabelian}(a) and (f).
In the opposite case $C_1\circ C_2$  the states are first swapped and then shifted to the neighboring unit cells.
The final states  in the two cases are different and they are shown in Fig.~\ref{fig:nonabelian}~(a,f) and Fig.~\ref{fig:nonabelian}~(e,l).
%The different final states confirm that $C_2\circ C_1\neq C_1\circ C_2$ -- and therefore the non-Abelian nature of the Thouless pumping in the system.

\begin{figure}[h!]
    \centering
    \includegraphics[width=\columnwidth]{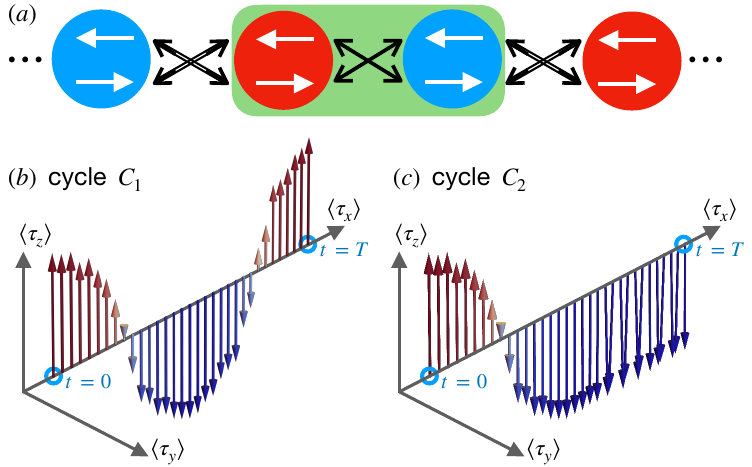}
    \caption{(a) Schematic representation of the spin Hamiltonian $H$ in Eq.~\eqref{eq:H_Yang} where the green box indicates a unit-cell. 
    The red and blue circles represent the pseudo-spin components $a_{n}$ and $b_{n}$, respectively object of the fields ${\bf B}_a$ and ${\bf B}_b$. 
    (b) Rotation of the pseudo-spin 
    $\tau=a$ and $\tau=b$ represented with upward red arrow and downward blue arrow respectively for the initial state $v_{a,n_0}^+$ along cycles $C_1$. 
    (c) Same as (b) for cycle $C_2$. 
    }
    \label{fig:spin_H}
\end{figure}

%{\it Yang monopole and chiral pumping} -- 
%Analogously to Eq.~\eqref{eq:H_k}, 
%we rewrite 
To  further elucidate its topological properties, it is useful to note that the ladder model given in Eq.~\eqref{eq:H} can be mapped onto a spinful Rice-Mele Hamiltonian with staggered magnetic field. %Hamiltonian for Yang monopoles. 
To this end, we perform a local unitary transformation $\mathcal{U}$ and redefine the chain in terms of spin $\sigma =\, \uparrow, \downarrow$ and pseudo-spin coordinates $\tau = a,b$ -- namely we set
$a_{k,\{ \uparrow, \downarrow\}}^{(\dagger)} = \pm e^{- i\frac{\pi}{4}} a_{k, U}^{(\dagger)} + e^{ i\frac{\pi}{4}}a_{k, D}^{(\dagger)} $ and
$b_{k,\{ \uparrow, \downarrow\}}^{(\dagger)} = e^{-  i\frac{\pi}{4}} a_{k, U}^{(\dagger)} \pm e^{ + i\frac{\pi}{4}}  a_{k, D}^{(\dagger)}$~\cite{Supple}. 
The Hamiltonian written in terms of the spinor creation and annihilation operators, 
$a_k^{(\dagger)}  = (a_{k,\uparrow}^{(\dagger)} ,a_{k, \downarrow}^{(\dagger)} )$ and $b_k^{(\dagger)}  = (b_{k,\uparrow}^{(\dagger)} ,b_{k, \downarrow}^{(\dagger)} )$ reads
\be
\begin{split}
H &= \sum_k \lf[ \lf( - J_k\, a_{k}^\dagger \sigma_y b_{k} + {\rm H.c. } \rg) + \rg.  \\
& \qquad  \lf. +\,   a_{k}^\dagger ({\mathbf B}_a\cdot \vec \sigma)\, a_{k}  + b_{k}^\dagger ({\mathbf B}_b\cdot \vec \sigma)\, b_{k} \rg]
\end{split}
\label{eq:H_Yang}
\ee
with ${\bf B}_a = ( 0 , - \mu, \rho )$ and ${\bf B}_b = (0, \mu , \rho )$. 
This system is visualized in Fig.~\ref{fig:spin_H}(a), where the pseudo-spin components $\tau = a,b $ are shown with red and blue circles respectively, and in it the spin degrees of freedom $\sigma =\,  \uparrow, \downarrow$ are represented with the white arrows. 

At  $t=0$ the Hamiltonian given in Eq.\eqref{eq:H_Yang}  is proportional to  $\sigma_z$ and the initial states are spinful particles localized on $a$ or $b$ sites, {\sl i.e.}
$v_{a,n_0}^\pm = |a_{n_0,\{ \uparrow, \downarrow\}}\ra$, and 
$v_{b,n_0}^\pm = |b_{n_0,\{ \uparrow, \downarrow\}}\ra$. 
We look at the evolution of these states by tracing the total density matrix $\hat{\rho}$ over the spacial  and spin indices respectively, $n$ and  $\sigma$, $\hat{\rho}_\tau = \text{Tr}_{n,\sigma} \hat{\rho}$. This reduced density matrix is then decomposed via the Pauli matrices to evaluate the expectation values of the pseudospin vector $\la \tau_i\ra  = \text{Tr}[\hat{\rho}_\tau \tau_i]$. 
Its three components are shown in Fig.~\ref{fig:spin_H}(b,c), where we show the evolution of $v_{a,n_0}^+$ along cycles $C_1$ and $C_2$.
%We note that at certain points along the cycles the evolution yields a state with maximally entangled  spin and pseudospin. At these points, the Bloch vector representing the pseudospin density matrix has a vanishing length. 
We note that at certain points along the cycles the Bloch vector representing the pseudospin density matrix has a vanishing length. This means that evolution can yield a state with maximally entangled spin and pseudospin.
%
%
%{\bf TO BE ADDED: discussion about Fig.~\ref{fig:spin_H}(b,c) concerning spin and charge density waves. }
%
Finally, we remark that the spinful Rice-mele model of Eq.\eqref{eq:H_Yang} can be related to the $SO(5)$ mean-field theory describing BCS and spin-density-wave (SDW) quasi-particles proposed in Ref.~\cite{demler1999}, the role of the latter being played by pseudospin excitations. The non-Abelian holonomy characterizing the Hamiltonian in Eq.~\eqref{eq:H_Yang} can be therefore related to a Yang monopole singularity. This is  analogous to what happens for the Zhang-Demler Hamiltonian~\cite{demler1999}.
Specifically,  Eq.~\eqref{eq:H_Yang} 
can be recast in terms of $\Phi_k^{(\dagger)}= (a_{k,\uparrow}^{(\dagger)},a_{k, \downarrow}^{(\dagger)},b_{k,\uparrow}^{(\dagger)},b_{k, \downarrow}^{(\dagger)})$ to
describe an $SO(5)$ spinor Hamiltonian $H_M =\sum_k \Phi_k ^\dagger  \lf[ \mathcal{B}_\mu \mathcal{L}_\mu   \rg] \Phi_k$  
where $\mathcal{L} = (L_1,L_2,L_3,L_4,L_5)$ are the Dirac matrices, and 
$\mathcal{B}=( - J_y , 0,-\mu,\rho,-J_x)$ is the correspondent $SO(5)$ field~\cite{Supple}.  
Note that in $\mathcal{B}$ the component corresponding to $L_2$ vanishes and it can be activated by {\it e.g.} turning into a complex variable the staggered interchain hopping $\rho e^{i \theta}$ in Eq.~\eqref{eq:H}. 
In this case,  in Eq.~\eqref{eq:H_Yang} we have ${\bf B}_a = ( \rho \sin\theta , - \mu, \rho \cos\theta )$ and ${\bf B}_b = (\rho \sin\theta, \mu , \rho\cos\theta )$, while the $SO(5)$ field in $H_M$ reads 
$\mathcal{B}=( - J_y , \rho \sin \theta,-\mu,\rho\cos \theta, -J_x)$. 
%This mapping hints that non-trivial holonomy arises from the non-zero overlap between the SDW and BCS quasiparticle states and the pump corresponds to the shift of such excitations. 

In conclusion, we have demonstrated how to realize non-Abelian Thouless pumping in a Rice-Mele ladder with time-dependent couplings. The model we propose exhibits doubly-degenerate Bloch bands and it has  both fundamental and practical significance.

First, it enables an exceptional degree of control over transport. By appropriately combining different pumping cycles, the proposed non-Abelian pumping protocol can (i) generate arbitrary lattice translations and (ii) implement all single-particle gates within the degenerate subspace. This result hold  potential for quantum computing and metrology, extending beyond the expectation values of observables and significantly enhancing the capabilities of standard holonomic gates ~\cite{zanardi1999,zhang2023geometric}.\\
Second, the Rice-Mele ladder systems discussed in this work can be implemented, not only in photonic setup~\cite{sun2022}, but also  using cold atoms in optical lattices~\cite{Atala2014} or quantum gas microscopy~\cite{greiner2009,bloch2010,gross2021,aidelsburger2024}. It may thus paves the way to the first experimental realization of non-Abelian Thouless pumping in a quantum many-body systems -- analogously to the Abelian Thouless pumping of interacting quantum particles in Rice-Mele chains~\cite{ke2017, nakagawa2018, liu2023, walter2023, viebahn2024}. \\
Third, we show that the Rice-Mele ladder can be related to an $SO(5)$ spinor model for Yang monopoles, 
hinting at a possible strategy to use Thouless pumps to investigate the dynamics of high-energy and strongly correlated systems.

{\it Acknowledgment --}  The authors acknowledge inspiring discussions with Monika Aidelsburger and Immanuel Bloch. V. B. and R.C. acknowledge financial support from PNRR MUR project PE0000023-NQSTI financed by the European Union – Next Generation EU.  This work was co-funded by European Union - PON Ricerca e Innovazione 2014-2020 FESR /FSC - Project ARS01$\_00734$ QUANCOM and PNRR MUR project CN 00000013-ICSC.

\bibliography{pumping,TOPO4}

\clearpage

\section*{Supplemental material:\\
Non-Abelian Thouless pumping in a Rice-Mele ladder }

\section{Bloch eigenstates}\label{app:eigenstates}

The $k$-space Hamiltonian of the proposed Rice-Mele ladder for $J_k = J_1 + J_2 e^{ik}$ reads 
\be 
\small
\begin{split}
\hspace{-10mm}
H &= \sum_{k}\sum_{M=U,D} \lf[J_k\, a_{k, M}^\dagger b_{k, M}+ {\rm H.c.}\rg]\\
&+ \mu  \sum_{k}  \lf[a_{k, U}^\dagger a_{k,U} - a_{k, D}^\dagger a_{k,D}   - b_{k, U}^\dagger b_{k,U} + b_{k, D}^\dagger b_{k,D}  \rg] \\
&+\rho  \sum_{k}  \lf[ a_{k, U}^\dagger a_{k,D}- b_{k, U}^\dagger b_{k,D} +{\rm H.c.}\rg] 
\end{split}
\label{eq:H_k0}
\ee 
Eq.~\eqref{eq:H_k0} can be rewritten for the four-dimensional spinor creation and annihilation operators as
$\Psi_k^{(\dagger)}=(a_{k,U}^{(\dagger)},a_{k,D}^{(\dagger)},b_{k,U}^{(\dagger)},b_{k,D}^{(\dagger)})$ and the Pauli matrices  $\sigma_j$ and $\tau_j$ 
%$\Psi_k=\left( |a_{k,U}\ra, |b_{k,U}\ra , |a_{k,D}\ra, |b_{k,D}\ra \right)$ 
\be 
\begin{split}
H= \sum_k \Psi_k^\dag \lf[  \lf(J_x \tau_x +J_y \tau_y\rg) \otimes\sigma_0 
+ \tau_z\otimes \lf(\rho \, \sigma_x+\mu\,\sigma_z \rg)\rg]\Psi_k
%H= \sum_k \Psi_k^\dag \lf[ \sigma_0\otimes \lf(J_x \tau_x +J_y \tau_y\rg)+ \lf(\rho \, \sigma_x+\mu\,\sigma_z \rg)\otimes\tau_z\rg]\Psi_k
\end{split}
\label{eq:H_k_app}
\ee
where $J_k= J_x + i J_y$  with $J_x=J_1+J_2\cos k$ and  $J_y=J_2 \sin k$. 
The eigenvalues of $H$ are
\be
\begin{split}
E_{\pm}(k)=\pm \sqrt{\mu^2 + \rho^2 + |J_k|^2}\equiv \pm \Delta
%E_\pm &= \pm \sqrt{\mu^2+\rho^2 + J_x^2 + J_y^2} 
\end{split}
\label{eq:Hk_evals}
\ee
\noindent
and the eigenvectors $\psi_{a}^{+},\psi_{b}^{+}$ associated to $E_+$ and $\psi_{a}^{-},\psi_{b}^{-}$ associated to $E_-$ read (see Eq.~\eqref{eq:H_k_app} of main text)
%\be
%\small
%\begin{split}
%v_1^\pm &= \left[ -J_x + i J_y, \mu \mp  \sqrt{\mu^2+\rho^2 + J_x^2 + J_y^2} , 0 , \rho \right] \\
%v_{2}^\pm  &=  \left[ \mu \pm  \sqrt{\mu^2+\rho^2 + J_x^2 + J_y^2} , J_x + i J_y ,  \rho , 0  \right] 
%\end{split}
%\label{eq:Hk_evectr1}
%\ee
%These eigenvectors do not form an orthogonal set -- {\it i.e.} in each band $\langle v_1^+,v_2^+\rangle\neq0$ and $\langle v_1^-,v_2^-\rangle\neq 0$. 
%Hence, we orthogonalize $v_2^\pm$ upon $v_1^\pm$ and normalized the eigenstates.  This leads to\be
%\small
%\begin{split}
%v_{1}^\pm &=  \frac{1}{\sqrt{2 \Delta ( \Delta \mp  \mu  )}} \left[ - J_k^*  |a_{k,U}\ra + (\mu \mp  \Delta )  |b_{k,U}\ra  + \rho  |b_{k,D}\ra\right] \\
%v_{2}^\pm &= \frac{1}{\sqrt{2 \Delta ( \Delta \mp \mu  )}}    \left[ \rho   |a_{k,U}\ra   - (\mu \mp \Delta ) |a_{k,D}\ra  +  J_k |b_{k,D}\ra \right] 
%\end{split}
%\label{eq:Hk_evectr8}
%\ee
%for $J_k = J_x + i J_y$ and $\Delta \equiv \sqrt{\mu^2+\rho^2 + J_x^2 + J_y^2}$.
%Hence, 
%
%for $\mathcal{R}_\pm = 1/\sqrt{2 \Delta ( \Delta \mp  \mu  )}$ the Bloch eigenstates of $H$ in Eq.~\eqref{eq:H_k_app} are 
\be
\small
\begin{split}
|\psi_{a}^{\pm}(k)\ra &= \frac{1}{\mathcal{R}_\pm} \lf[\rho |a_{k,U}\ra + (-\mu \pm \Delta ) |a_{k,D}\ra+J_k
 |b_{k,D}\ra\rg]\\ 
|\psi_{b}^{\pm}(k)\ra &=  \frac{1}{\mathcal{R}_\pm} \lf[ - J_k^* |a_{k,U}\ra + (\mu \mp  \Delta ) |b_{k,U}\ra + \rho |b_{k,D}\ra\rg] 
\end{split}
\label{eq:psi1_app}
\ee
with $\mathcal{R}_\pm = 1/\sqrt{2 \Delta ( \Delta \mp  \mu  )}$.
Note that the above eigenvectors are orthonormalized.

\section{Holonomies $\Gamma^\pm_{t}$ }\label{app:holo_0}

We now calculate the holonomy matrices for $\alpha=\pm$ 
\be
[\Gamma^\alpha_{t}]_{ij}=\la \psi_{i}^{\alpha}(k)  | \partial_t |\psi_{j}^{\alpha}(k)\ra \label{eq:holo1_app}
\ee
with $\alpha=\pm$.
%defined for the eigenvectors $|\psi_{a}^{\pm}(k)\ra,|\psi_{b}^{\pm}(k)\ra$ in Eq.~\eqref{eq:psi1_app}. 
%
Each element of $\Gamma^\alpha_{t}$ in Eq.~\eqref{eq:holo1_app} is calculated as
\be
[\Gamma^\alpha_{t}]_{ij}= {\bf A}_{i,j}^\alpha \cdot {\bf V}
\ee
where  ${\bf V}=(\dot{J}_x,\dot{J}_y,\dot{\rho},\dot{\mu})$ and ${\bf A}^\alpha_{i,j}$ are given by
%Eq.~\eqref{eq:holo1_app}, 
%we split the chain rule in the time derivative $\partial_t |\psi_{j}^{\pm}(k)\ra$ by first computing the auxiliary four-dimensional vectors $A_{i,j}^\pm$  defined as
 %and  ${\bf V}=(\dot{J}_x,\dot{J}_y,\dot{\rho},\dot{\mu})$.
\be\label{Aaux}
\begin{split}
A_{i,j}^\alpha &= i \left\{  \langle \psi_{i}^{\alpha}(k) | \partial_{J_x} |\psi_{j}^{\alpha}(k) \rangle, \langle \psi_{i}^{\alpha} (k)| \partial_{J_y} | \psi_{j}^{\alpha} (k) \rangle , \right. \\ 
&\left.\qquad \langle \psi_{i}^{\alpha}(k)|\partial_{\rho}| \psi_{j}^{\alpha} (k) \rangle , \langle \psi_{i}^{\alpha}(k)| \partial_{\mu} |\psi_{j}^{\alpha} (k)\rangle \right\}. 
\end{split}
\ee
Starting from the above equation, using the eigenstates given in \eqref{eq:psi1_app} we get
%Then $\left[\Gamma^\pm_{t} \right]_{i,j} =  A^\mu_{i,j}$%\cdot | \upsilon \rangle$ where $\upsilon=(\dot{J}_x,\dot{J}_y,\dot{\rho},\dot{\mu})$ is the vector of the time-derivatives of the system's parameters. 
%In our case, the four auxiliary vectors are
\bea
\footnotesize
A_{a,a}^{\alpha} 
%= i \left(  \langle v_1, \partial_{J_x} v_1\rangle, \langle v_1, \partial_{J_y} v_1\rangle , \langle v_1, \partial_{\rho} v_1\rangle , 
%  \langle v_1, \partial_{\mu} v_1\rangle \right)  \\
%&\hspace{-43.5mm} 
&=&   \frac{1}{\mathcal{R}_\pm^2} \left( - J_y, J_x, \rho, 0 \right)    \\ 
A_{a,b}^{\alpha} 
%= i \left(  \langle v_1, \partial_{J_x} v_2\rangle, \langle v_1, \partial_{J_y} v_2\rangle , \langle v_1, \partial_{\rho} v_2\rangle , \langle v_1, \partial_{\mu} v_2\rangle \right)  \notag \\
%& \hspace{-30.5mm} 
&=&   \frac{1}{\mathcal{R}_\pm^2} \left( i\rho, - \rho, i (J_x + i J_y),  0 \right) \notag   \\ 
A_{b,a}^{\alpha} 
%= i \left(  \langle v_2, \partial_{J_x} v_1\rangle, \langle v_2, \partial_{J_y} v_1\rangle , \langle v_2, \partial_{\rho} v_1\rangle , \langle v_2, \partial_{\mu} v_1\rangle \right) \notag  \\
%&\hspace{-28mm} 
&=&   \frac{1}{\mathcal{R}_\pm^2} \left( -i \rho,  - \rho, i (J_x - i J_y),  0 \right) \notag  \\
A_{b,b}^{\alpha} 
%= i \left(  \langle v_2, \partial_{J_x} v_2\rangle, \langle v_2, \partial_{J_y} v_2\rangle , \langle v_2, \partial_{\rho} v_2\rangle , \langle v_2, \partial_{\mu} v_2\rangle \right) \notag  \\
%&\hspace{-43mm} 
&=&   \ \frac{1}{\mathcal{R}_\pm^2} \left(  J_y, - J_x, 0, 0 \right) \notag 
\label{eq:Gaux}
\eea
Eventually, recalling that $J_x = J_1+J_2\cos k$, $ J_y = J_2 \sin k$, in terms of the Pauli matrices $\sigma_j$ 
we recover Eq.(8) of main text:
\be
\begin{split}
\Gamma^\pm_{t}&=
\frac{1}{\mathcal{R}_\pm^2}
\lf[\lf(J_2 \Dot{J_1}-J_1\Dot{J_2}\rg) \sin k\,\sigma_z  
+(\Dot{J_1}\rho-J_1\Dot{\rho}) \sigma_y \rg.\\
&\qquad\quad \lf. + (\Dot{J_2}\rho-J_2\Dot{\rho})( \cos k\, \sigma_y - \sin k\, \sigma_x)  \rg]
\end{split}
\label{eq:holoZ1_app}
\ee

\section{Adiabatic conditions}\label{app:adiabatic}

\begin{figure}[h!]
    \centering
    \includegraphics[width=0.825\columnwidth]{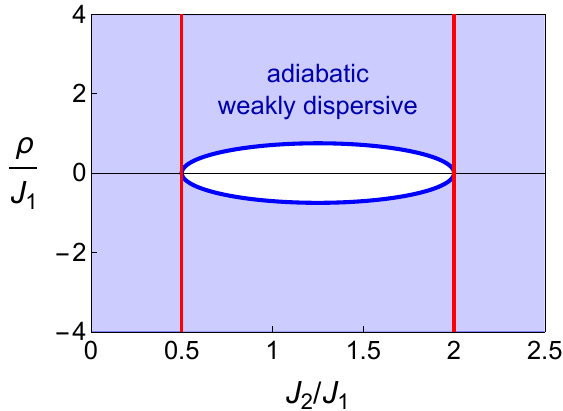}
    \caption{Diagram of the adiabatic weakly dispersive regime as function of $\frac{J_1}{\rho}$ versus $\frac{J_2}{J_1}$ for $\mu=0$. The two blue curves are the bounds $g_\alpha$ in Eq.~\eqref{eq:Acond12_app}, while the red vertical lines indicate $\frac{J_2}{J_1} = 0.5$ and $\frac{J_2}{J_1} = 2$ respectively.}
    \label{fig:pumping_cycle}
\end{figure}

The conditions for adiabatic and weakly dispersive pumping  -- here recalled (see also Refs.\cite{hu2019,liu2024} for more details)
\bea
\label{eq:Acond1_app}
&& \min_{k,t} |E_{+}(k) - E_{-}(k)| T \gg 1 \\
\label{eq:Acond2_app}
&& \hspace{4.5mm}  \max_{k,t}\lf| \partial_k E_{\pm}(k) \rg|  T   \ll a
\eea
For lattice spacing set as $a=1$,  Eqs.~(\ref{eq:Acond1_app},\ref{eq:Acond2_app}) imply that a period $T$ has to be chosen within 
\bea
\label{eq:Acond3_app}
\max_{k,t} \frac{1}{ |E_{+}(k) - E_{-}(k)| }  \ll T \ll \min_{k,t}\frac{1}{\lf| \partial_k E_{\pm}(k) \rg|}
\eea
The band gap $|E_{+}(k) - E_{-}(k)|$ has a minimum at $k=\pi$ for each $t\in[0,T]$. Hence, the maximum of the inverse of the bandgap is 
\be
\begin{split}
\max_{k,t} \frac{1}{ |E_{+}(k) - E_{-}(k)| } = \frac{1}{2\sqrt{\mu^2+\rho^2+ (J_1  - J_2 )^2}}
\end{split}
\ee
On the other hand, the upper bound of the derivative $\partial_k E_{\pm}(k)$ can be approximated as 
\be
\lf| \partial_k E_{\pm}(k) \rg| \lesssim \frac{J_1 J_2 }{\sqrt{\mu^2+\rho^2+(J_1-J_2)^2 }}
\ee 
Eq.~\eqref{eq:Acond3_app} then reads as
\bea
\label{eq:Acond4_app}
&\frac{1}{2\sqrt{\mu^2+\rho^2+ (J_1  - J_2 )^2}}  \ll T \ll \frac{\sqrt{\mu^2+\rho^2+(J_1-J_2)^2}}{J_1 J_2 }
\eea
Let us observe that the inequality 
\be
\label{eq:Acond5_app}
\frac{1}{2\sqrt{\mu^2+\rho^2+ (J_1  - J_2 )^2}}  \leq\frac{\sqrt{\mu^2+\rho^2+(J_1-J_2)^2 }}{J_1 J_2 } 
\ee
does not hold for any parameter choice, implying that both inequalities in Eq.~\eqref{eq:Acond3_app} can not be simultaneously satisfied. 
%\noindent
For $J_2= \alpha J_1\equiv \alpha J$ -- {\it i.e.} $\alpha = \frac{J_2}{J_1}$ -- Eq.~\eqref{eq:Acond5_app} reduces to 
\be
\small
\begin{split}
\label{eq:Acond8_app}
%&\frac{1}{2\sqrt{\mu^2+\rho^2+ J^2(1  - \alpha )^2}}  \leq\frac{\sqrt{\mu^2+\rho^2+J^2 (1+\alpha^2) }}{\alpha J^2 }  \\
%\Leftrightarrow\quad  &
%\frac{1}{4 [\mu^2+\rho^2+ J^2(1  - \alpha )^2]}  \leq\frac{\mu^2+ \rho^2+J^2 (1+\alpha^2) }{\alpha^2 J^4 } \\
%\Leftrightarrow\quad  &
%\alpha^2 J^4  \leq 4 [\mu^2+ \rho^2+ J^2(1  - \alpha )^2] [\mu^2+ \rho^2+J^2 (1+\alpha^2) ] \\
%\Leftrightarrow\quad  &
%\alpha^2 J^4  \leq 4 (\mu^2+\rho^2)^2 + 8 (\mu^2+\rho^2) J^2 (1  - \alpha + \alpha^2 ) \\
%&\qquad\quad + 4 J^4 (1+\alpha^2)(1  - \alpha )^2 \\
%\Leftrightarrow\quad  &
%[ \alpha^2 - 4 (1+\alpha^2)(1  - \alpha )^2 ] J^4  \\
%& - 8 (\mu^2+\rho^2) (1  - \alpha + \alpha^2 )  J^2  - 4 (\mu^2+\rho^2)^2  \leq  0 \\
%\Leftrightarrow\quad  & 
p_2(\alpha) J^4 - 8 (\mu^2+\rho^2) p_1(\alpha)  J^2  - 4 (\mu^2+\rho^2)^2  \leq  0
\end{split}
\ee
for $p_2(\alpha) =  \alpha^2 - 4 (1  - \alpha )^4$ and $p_1(\alpha) = (1  - \alpha )^2$. 
On the one hand $p_1(\alpha)\geq 0$ for every $\alpha$. On the other hand $p_2$ posses two real roots at $\alpha_0 =\frac{1}{2}$ and $\alpha_1 =2$. 
For $0\leq \alpha\leq \alpha_0$ and $ \alpha > \alpha_1$ then $p_2(\alpha) <0 $ and the condition Eq.~\eqref{eq:Acond5_app} is always satisfied. For $\alpha_0 < \alpha < \alpha_1$ then $p_2(\alpha) > 0$ and Eq.~\eqref{eq:Acond8_app} yields the range
\be
\begin{split}
\label{eq:Acond12_app}
&- g_\alpha\sqrt{\mu^2+\rho^2} \leq J  \leq   g_\alpha \sqrt{\mu^2+\rho^2}
\end{split}
\ee
defined for $g_\alpha \equiv \sqrt{ \frac{4 p_1 + 2 \sqrt{4 p_1^2 + p_2 }}{ p_2}}$. 
This existence condition of the adiabatic weakly dispersive pumping regime for $\mu=0$ is shown in Fig.~\ref{fig:pumping_cycle}. 

\section{Wilson loops of cycles $C_1$ and $C_2$}\label{app:holo}

Let us compute the path integral 
\be
 \mathcal{P} \exp\left[ i\int_{t_0}^{t_0+T} \Gamma^\pm_{t} dt \right] .
\label{eq:wilson_app}
\ee
of the Wilson loop $W^\pm (t_0,t_0+T)$ associated to the holonomy $\Gamma^\pm_{t}$ in Eq.~\eqref{eq:holoZ1_app} for both cycles $C_1$ and $C_2$.

\subsection{Cycle $C_1$}\label{app:holoC1}

\begin{figure}[h!]
    \centering
    \includegraphics[width=0.25\textwidth]{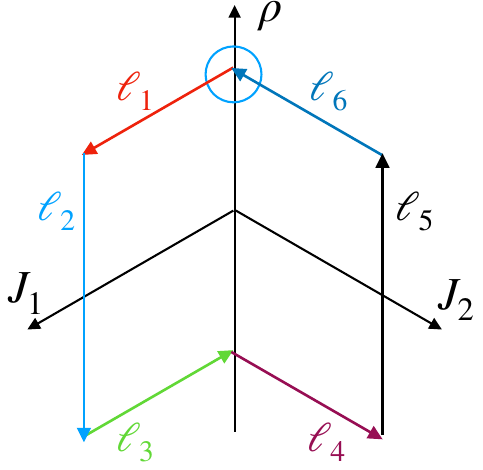}
    \caption{(a) Schematic representation of the pumping cycle $\mathcal{C}_1$ split in six bits $\{\ell_1,\dots,\ell_6\}$ shown in different colors. 
    }
    \label{fig:cycle_C1_split}
\end{figure}

Let us split the cycle $C_1$ in six segments $\{\ell_1,\dots,\ell_6\}$ as shown in Fig.~\ref{fig:cycle_C1_split} and rewrite $\Gamma^+_{t}$ in Eq.~\eqref{eq:holoZ1_app} along each segment for $\Sigma_k = \cos k\, \sigma_y - \sin k\, \sigma_x$. 

\noindent
This yields six terms
\be
\begin{split}
&\ell_1:
\begin{cases}
\rho=1 \\
J_1 = \theta\\
J_2=0
\end{cases}
 \Rightarrow\quad 
\Gamma^+_{t}  
=  \frac{\Dot{J_1}\rho-J_1\Dot{\rho}}{2 (\rho^2+  J_1^2)}  \sigma_y
%= \frac{1 \cdot 1- \theta \cdot 0}{2 (1+  \theta^2)}  \sigma_y
=   \frac{1}{2 (1 + \theta^2)} \sigma_y 
 \end{split}
\notag %  \label{eq:holonomy_t_split1}
\ee
 \be
\begin{split}
& \ell_2:
\begin{cases}
\rho=-\theta \\
J_1 = 1\\
J_2=0
\end{cases}
\Rightarrow\quad 
\Gamma^+_{t} 
=  \frac{\Dot{J_1}\rho-J_1\Dot{\rho}}{2 (\rho^2+  J_1^2)}  \sigma_y
%= \frac{0\cdot (-\theta) - 1\cdot  (-1)}{2 (1+  \theta^2)}  \sigma_y
 =  \frac{1}{2 (1 + \theta^2)} \sigma_y  
 \end{split}
\notag % \label{eq:holonomy_t_split2}
\ee
 \be
\begin{split}
& \ell_3:
\begin{cases}
\rho=- 1 \\
J_1 = - \theta\\
J_2=0
\end{cases}
 \Rightarrow\quad 
\Gamma^+_{t} 
=  \frac{\Dot{J_1}\rho-J_1\Dot{\rho}}{2 (\rho^2+  J_1^2)}  \sigma_y
%= \frac{(-1) \cdot (-1) - (-\theta) \cdot 0 }{2 (1+  \theta^2)}  \sigma_y
 =  \frac{ 1}{2 (1 + \theta^2)} \sigma_y  
\end{split}
\notag %  \label{eq:holonomy_t_split3}
\ee
\be
\begin{split}
 &\ell_4:
\begin{cases}
\rho=- 1 \\
J_1 = 0\\
J_2= \theta
\end{cases}
 \Rightarrow\quad \ 
\Gamma^+_{t} 
=  \frac{\Dot{J_2}\rho-J_2\Dot{\rho}}{2 (\rho^2+  J_2^2)}  \Sigma_k 
% =  \frac{1 \cdot (-1) - \theta \cdot 0}{2 (1 + \theta^2)} \Sigma_k 
  =  \frac{-1}{2 (1 + \theta^2)} \Sigma_k  
\end{split}
\notag %  \label{eq:holonomy_t_split4}
\ee
\be
\begin{split}
  &\ell_5:
\begin{cases}
\rho=\theta \\
J_1 = 0\\
J_2= 1
\end{cases}
\Rightarrow\quad \ 
\Gamma^+_{t} 
=  \frac{\Dot{J_2}\rho-J_2\Dot{\rho}}{2 (\rho^2+  J_2^2)}  \Sigma_k 
% =  \frac{0 \cdot \theta - 1 \cdot 1 }{2 (1 + \theta^2)}   \Sigma_k   
   =  \frac{-1}{2 (1 + \theta^2)} \Sigma_k 
\end{split}
\notag %  \label{eq:holonomy_t_split5}
\ee
\be
\begin{split}
  &\ell_6:
\begin{cases}
\rho= 1 \\
J_1 = 0\\
J_2=- \theta
\end{cases}
\Rightarrow\quad \ 
\Gamma^+_{t} 
=  \frac{\Dot{J_2}\rho-J_2\Dot{\rho}}{2 (\rho^2+  J_2^2)}  \Sigma_k 
% =  \frac{(-1)\cdot 1- (-\theta) \cdot 0 }{2 (1 + \theta^2)}  \Sigma_k 
    =  \frac{-1}{2 (1 + \theta^2)} \Sigma_k
\end{split}
\label{eq:holonomy_t_split6}
\ee
These yield six integrals
\be
\begin{split}
% \oint_{\mathcal{C}_1} \Gamma^+_{t} d\theta = \sum_{s=1}^6  \oint_{\ell_s} \Gamma^+_{t} d\theta
% \qquad\text{with}\qquad
 &\oint_{\ell_s} \Gamma^+_{t} d\theta = \frac{\sigma_y }{2} \int_0^1  \frac{1}{ 1 + \theta^2} d\theta 
% = \frac{1}{2} \frac{\pi}{4}  \sigma_y  
 = \frac{\pi}{8}  \sigma_y \quad\text{for $s=1,3$} \\
  &\oint_{\ell_s} \Gamma^+_{t} d\theta = \pm \frac{\sigma_y }{2} \int_{-1}^1  \frac{1}{ 1 + \theta^2} d\theta 
 %= \pm \frac{1}{2} \frac{\pi}{2}  \sigma_y  
 =  \pm \frac{\pi}{4}  \sigma_y \quad\text{for $s=2,5$} \\
 &\oint_{\ell_s} \Gamma^+_{t} d\theta =-  \frac{\Sigma_k}{2} \int_0^1  \frac{1}{ 1 + \theta^2} d\theta 
 %= - \frac{1}{2} \frac{\pi}{4} \Sigma_k  
 = - \frac{\pi}{8}  \Sigma_k\quad\text{for $s=4,6$} \\
 \end{split}
\label{eq:int1}
\ee
which result in the path integral in Eq.~\eqref{eq:wilson_app}, and ultimately in the Wilson loop of the cycle $C_1$, up to for a dynamical phase factor
\be
W^\pm_{C_1} =%(0,T) =
 \begin{pmatrix}
 e^{ik} & 0\\
0  & e^{-ik}
 \end{pmatrix} 
\label{eq:W_C1_app}
\ee

\subsection{Cycle $C_2$}\label{app:holoC2}

\begin{figure}[h!]
    \centering
    \includegraphics[width=0.25\textwidth]{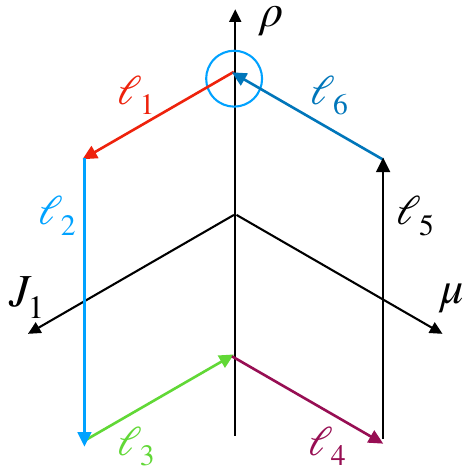}
    \caption{(a) Schematic representation of the pumping cycle $\mathcal{C}_2$ split in six bits $\{\ell_1,\dots,\ell_6\}$ shown in different colors. 
    }
    \label{fig:cycle_C2_split}
\end{figure}

Let us split the cycle $C_2$ in six segments $\{\ell_1,\dots,\ell_6\}$ as shown in Fig.~\ref{fig:cycle_C2_split} and rewrite $\Gamma^+_{t}$ in Eq.~\eqref{eq:holoZ1_app} along each segment. 
%
%\noindent
The first three segments $\ell_1,\ell_2,\ell_3$ are the same as in cycle $C_1$. The latter ones instead yield
\be
\begin{split}
 &\ell_4:
\begin{cases}
\rho=- 1 \\
J_1 = 0\\
\mu= \theta
\end{cases}
\quad \Rightarrow\quad \ 
\Gamma^+_{t} 
=  \frac{\Dot{J_1}\rho-J_1\Dot{\rho}}{2 (\rho^2 + \mu^2+  J_1^2)}  \sigma_y
% =  \frac{0  \cdot (-1) -  0 \cdot 0}{2 (1 + \theta^2)} \sigma_y
  = 0
\end{split}
\notag %\label{eq:holonomy_t_split4}
\ee
\be
\begin{split}
  &\ell_5:
\begin{cases}
\rho=\theta \\
J_1 = 0\\
\mu= 1
\end{cases}
\quad \Rightarrow\quad \ 
\Gamma^+_{t} 
=   \frac{\Dot{J_1}\rho-J_1\Dot{\rho}}{2 (\rho^2 + \mu^2+  J_1^2)}  \sigma_y
% =  \frac{0 \cdot \theta - 1 \cdot 0 }{2 (1 + \theta^2)}   \sigma_y   
   = 0
\end{split}
\notag %\label{eq:holonomy_t_split5}
\ee
\be
\begin{split}
  &\ell_6:
\begin{cases}
\rho= 1 \\
J_1 = 0\\
\mu=- \theta
\end{cases}
\quad \Rightarrow\quad \ 
\Gamma^+_{t} 
=  \frac{\Dot{J_1}\rho-J_1\Dot{\rho}}{2 (\rho^2 + \mu^2+  J_1^2)}  \sigma_y
% =  \frac{0 \cdot 1- 0  \cdot 0 }{2 (1 + \theta^2)}  \sigma_y 
    =  0
\end{split}
\label{eq:holonomy_t_split6_c2}
\ee
This results in trivial integrals over 
\be
\begin{split}
 &\oint_{\ell_s} \Gamma^+_{t} d\theta = 0 \qquad\text{for $s=4,5,6$} 
 \end{split}
\label{eq:int2}
\ee
which ultimately results the Wilson loop of the cycle $C_2$ to be (up to for a dynamical phase factor)
\be
W^\pm_{C_2} =%(0,T) = 
 \begin{pmatrix}
 0 & 1\\
-1  & 0 
 \end{pmatrix} 
\label{eq:W_C2_app}
\ee

\section{One period evolution}\label{app:1period}

The initial states $v_{a,n_0}^\pm, v_{b,n_0}^\pm$ at  $J_1=J_2=\mu=0$ and $\rho=\rho_0\neq0$ in real space read
\bea
\label{eq:IC1a_app}
v_{a,n_0}^\pm  &=& \frac{\delta_{n,n_0}}{\sqrt{2}} \lf[ |a_{n,D} \ra \pm |a_{n,U}\ra  \right]  \\
\label{eq:IC1b_app}
v_{b,n_0}^\pm  &=& \frac{\delta_{n,n_0}}{\sqrt{2}} \lf[ |b_{n,D} \ra \mp |b_{n,U}\ra  \right].
\eea
while in momentum space $|\psi_{n_0}^{\pm}(t_0)\rangle = \sum_{k,\nu} c_{\nu} |\psi_{\nu}^{\pm}(k)\ra e^{i k n_0}$ turn to
\bea
\label{eq:IC1a_Bs_app}
v_{a,n_0}^\pm  
&=& \sum_{k,\nu} \left[ 1 |\psi_{a}^{\pm}(k)\ra + 0  |\psi_{b}^{\pm}(k)\right]  e^{i k n_0} \\
&=&  \sum_{k,\nu} |\psi_{a}^{\pm}(k)\ra e^{i k n_0}  \notag \\
\label{eq:IC1b_Bs_app}
v_{b,n_0}^\pm  
&=& \sum_{k,\nu} \left[ 0 |\psi_{a}^{\pm}(k)\ra + 1  |\psi_{b}^{\pm}(k)\right]  e^{i k n_0} \\
&=& \sum_{k,\nu} |\psi_{b}^{\pm}(k)\ra e^{i k n_0}  \notag
\eea
The propagation in the adiabatic regime is dictated by the Wilson loop  $W$
\be
|\psi_{n_0}^{\pm}(t_0+T)\rangle 
= 
\sum_{k\nu\eta}
 c_{\nu }\left[ W^\pm(t_0,t_0+T)\right]_{\eta\nu}|\psi_{\eta}^{\pm}(k)\ra  e^{i k n_0} 
\label{eq:propagation1_app}
\ee
In the case of cycles $C_1$ with loop in Eq.~\eqref{eq:W_C1_app}, the states $v_{a,n_0}^\pm, v_{b,n_0}^\pm$ at $t=0$ are mapped to $\bar{v}_{a,n_0}^\pm,\bar{v}_{b,n_0}^\pm$ at $t=T$
\bea
\label{eq:IC1a_Bs_app}
\bar{v}_{a,n_0}^\pm
&=& \sum_{k,\nu} 1 \left[  e^{i k } |\psi_{a}^{\pm}(k)\ra + 0  |\psi_{b}^{\pm}(k)\right]  e^{i k n_0} \\
&=&  \sum_{k,\nu}   e^{i k } |\psi_{a}^{\pm}(k)\ra e^{i k n_0}  
%\notag \\
%&=& \sum_{k,\nu}  |\psi_{a}^{\pm}(k)\ra e^{i k (n_0+1)} 
= v_{a,n_0+1}^\pm  \notag \\
\label{eq:IC1b_Bs_app}
\bar{v}_{b,n_0}^\pm 
&=& \sum_{k,\nu} 1 \left[ 0 |\psi_{a}^{\pm}(k)\ra + e^{ - i k } |\psi_{b}^{\pm}(k)\right]  e^{i k n_0} \\
&=& \sum_{k,\nu} e^{ - i k } |\psi_{b}^{\pm}(k)\ra e^{i k n_0} 
%\notag \\
%&=& \sum_{k,\nu}  |\psi_{b}^{\pm}(k)\ra e^{i k (n_0-1)} 
= v_{b,n_0-1}^\pm  \notag
\eea
Hence
\be
 C_1: 
\begin{cases}
v_{a,n_0}^\pm  \longmapsto v_{a,n_0+1}^\pm  \\
v_{b,n_0}^\pm \longmapsto v_{b,n_0-1}^\pm 
\end{cases}
\label{eq:C1_app}
\ee
Instead, in the case of cycles $C_2$ with loop in Eq.~\eqref{eq:W_C2_app}, it follows that 
\bea
\label{eq:IC2a_Bs_app}
\bar{v}_{a,n_0}^\pm
&=& \sum_{k,\nu} 1 \left[  0 |\psi_{a}^{\pm}(k)\ra + 1  |\psi_{b}^{\pm}(k)\right]  e^{i k n_0} \\
&=&  \sum_{k,\nu}   |\psi_{b}^{\pm}(k)\ra e^{i k n_0} = v_{b,n_0}^\pm \notag \\
\label{eq:IC2b_Bs_app}
\bar{v}_{b,n_0}^\pm 
&=& \sum_{k,\nu} 1 \left[ (-1) |\psi_{a}^{\pm}(k)\ra +0|\psi_{b}^{\pm}(k)\right]  e^{i k n_0} \\
&=& -\sum_{k,\nu}  |\psi_{a}^{\pm}(k)\ra e^{i k n_0} =-  v_{a,n_0}^\pm   \notag
\eea
Hence
\be
C_2:
\begin{cases}
v_{a,n_0}^\pm  \longmapsto v_{b,n_0}^\pm  \\
v_{b,n_0}^\pm  \longmapsto -  v_{a,n_0}^\pm  
\end{cases}
\label{eq:C2_app}
\ee

\section{Mapping to an SO(5) model}

Let us consider a unitary transformation defined by a $4\times 4$ unitary matrix $\mathcal{U}$ composed of a gauge transformation $\mathcal{G}$ and a coordinate rotation $ \mathcal{T}$
\be
\mathcal{U} = \mathcal{T} \mathcal{G} 
=
%\underbrace{
 \frac{1}{\sqrt{2}} 
\begin{pmatrix}
1 & -1 & 0 & 0 \\
1 & 1 & 0 & 0 \\
0 & 0 & 1 & -1 \\
0 & 0 & 1 & 1 \\
\end{pmatrix} 
%}_{\textcolor{blue}{rotation}}
%\underbrace{
\begin{pmatrix}
1 & 0 & 0 & 0 \\
0 & 1 & 0 & 0 \\
0 & 0 & 1 & 0 \\
0 & 0 & 0 & -1 \\
\end{pmatrix} 
\label{eq:U_tr}
\ee
The Bloch Hamiltonian of the proposed Rice-Mele ladder in Eq.~\eqref{eq:H_k0} in these new coordinates reads
\be 
\small
\begin{split}
H&= J_k \sum_k  \lf[\, a_{k,U}^{\dagger} b_{k,D} + a_{k,D}^{\dagger} b_{k,U} + {\rm H.c.} \rg]\\
%&+ J_2\sum_k  \lf[a_{n,U}^{\dagger} b_{n-1,D} + a_{n,D}^{\dagger} b_{n-1,U} + {\rm H.c.} \rg] \\
&+ \rho %\cos\theta
\sum_k  \lf[- a_{k,U}^{\dagger} a_{k,U} + a_{k,D}^{\dagger} a_{k,D} - b_{k,U}^{\dagger} b_{k,U}+  b_{k,D}^{\dagger} b_{k,D}   \rg] \\
&+ \mu \sum_k  \lf[a_{n,U}^{\dagger} a_{k,D} - b_{k,U}^{\dagger} b_{k,D} + {\rm H.c.} \rg]  
\end{split} 
\label{eq:H_app2}
\ee
where we recall that $J_k = J_1 + J_2 e^{ik}$. 
Via the Pauli matrices $\vec \sigma = (\sigma_x,\sigma_y,\sigma_z)$ and the fields  
${\bf B}_a = (\mu, 0 , \rho )$,   ${\bf B}_b = (-\mu, 0, \rho )$, 
%${\bf B}_a = (\mu, \rho\sin\theta , \rho \cos\theta )$,   ${\bf B}_b = (-\mu,  \rho\sin\theta , \rho \cos\theta)$, 
this Hamiltonian can be recast as 
\be 
\small
\begin{split}
H&= \sum_k \lf[J_k\, 
\begin{pmatrix}
a_{k,U}^{\dagger} , 
a_{k,D}^{\dagger} 
\end{pmatrix} 
\sigma_x 
\begin{pmatrix}
b_{k,U} \\
b_{k,D}
\end{pmatrix} 
+ {\rm H.c.}\rg]\\
&+\sum_{k} \lf[
\begin{pmatrix}
a_{k,U}^{\dagger} , 
a_{k,D}^{\dagger} 
\end{pmatrix}  ({\bf B}_a\cdot \vec \sigma) 
\begin{pmatrix}
a_{k,U} \\
a_{k,D}
\end{pmatrix}  \rg. \\
& \lf. \qquad +
\begin{pmatrix}
b_{k,U}^{\dagger} , 
b_{k,D}^{\dagger} 
\end{pmatrix}  ({\bf B}_b\cdot \vec \sigma) 
\begin{pmatrix}
b_{k,U} \\
b_{k,D}
\end{pmatrix}\rg] 
\end{split} 
\label{eq:H_app3}
\ee
We then further re-orient the spin coordinates by rotating $\sigma_y \longmapsto \sigma_x$, $\sigma_x \longmapsto - \sigma_y$ while leaving $\sigma_z$ and the pseudo-spin coordinates $\tau_j$ untouched.
The composition of these three coordinates transformations (namely, $\mathcal{T}$ and  $\mathcal{G}$ in Eq.~\eqref{eq:U_tr} and this axis re-orientation) results in the coordinates 
$a_{k,\{ \uparrow, \downarrow\}}^{(\dagger)} = \pm e^{- i\frac{\pi}{4}} a_{k, U}^{(\dagger)} + e^{ i\frac{\pi}{4}}a_{k, D}^{(\dagger)} $ and
$b_{k,\{ \uparrow, \downarrow\}}^{(\dagger)} = e^{-  i\frac{\pi}{4}} a_{k, U}^{(\dagger)} \pm e^{ + i\frac{\pi}{4}}  a_{k, D}^{(\dagger)}$. 
The Hamiltonian in Eq.~\eqref{eq:H_app3} written in terms of the spinor creation and annihilation operators 
$a_k^{(\dagger)}  = (a_{k,\uparrow}^{(\dagger)} ,a_{k, \downarrow}^{(\dagger)} )$ and $b_k^{(\dagger)}  = (b_{k,\uparrow}^{(\dagger)} ,b_{k, \downarrow}^{(\dagger)} )$ reads
\be
\begin{split}
H &= \sum_k \lf[ \lf( - J_k\, a_{k}^\dagger \sigma_y b_{k} + {\rm H.c. } \rg) + \rg.  \\
& \qquad  \lf. +\,   a_{k}^\dagger ({\mathbf B}_a\cdot \vec \sigma)\, a_{k}  + b_{k}^\dagger ({\mathbf B}_b\cdot \vec \sigma)\, b_{k} \rg]
\end{split}
\label{eq:H_Yang_app}
\ee
with ${\bf B}_a = ( 0 , - \mu, \rho )$ and ${\bf B}_b = (0, \mu , \rho )$. 
In terms of a four-dimensional spinor $\Phi_k^{(\dagger)}= (a_{k,\uparrow}^{(\dagger)},a_{k, \downarrow}^{(\dagger)},b_{k,\uparrow}^{(\dagger)},b_{k, \downarrow}^{(\dagger)})$, Eq.~\eqref{eq:H_Yang_app} can be rewritten as 
\be
\begin{split}
H &= \sum_k \lf[- J_x\, \Phi_k ^\dagger (\tau_x\otimes\sigma_y) \Phi_k  -
J_y\, \Phi_k ^\dagger (\tau_y\otimes\sigma_y) \Phi_k  \rg] \\
&+\sum_{n} \lf[ 
-\mu\, \Phi_k ^\dagger (\tau_z\otimes\sigma_y) \Phi_k  + 
\rho\, \Phi_k ^\dagger (\tau_0\otimes\sigma_z) \Phi_k
\rg]
\end{split}
\label{eq:H_k_app3}
\ee
The Hamiltonian Eq.~\eqref{eq:H_k_app3} expressed by the Dirac matrices of Yang monopoles 
$L_1 = \tau_y\otimes\sigma_y$, 
$L_2 = \tau_0\otimes\sigma_x$,
$L_3 = \tau_z\otimes\sigma_y $,
$L_4 = \tau_0\otimes\sigma_z$ and 
$L_5 = \tau_x\otimes\sigma_y$, 
reads
\be
\begin{split}
H_M 
&=\sum_k \Phi_k ^\dagger  \lf[ 
 - J_y L_1  
-\mu L_3  + 
\rho  L_4 
 -J_x L_5
 \rg] \Phi_k \\
  &=\sum_k \Phi_k ^\dagger  \lf[ 
 \mathcal{B}_\mu \mathcal{L}_\mu  
 \rg] \Phi_k 
\end{split}
\label{eq:H_Yang_v3}
\ee 
for $\mathcal{B}=( - J_y , 0,-\mu,\rho,-J_x)$ and $\mathcal{L} = (L_1,L_2,L_3,L_4,L_5)$. 
Let us turn the transversal hopping in Eq.~\eqref{eq:H_k0} complex $ \rho e^{i\theta}$. 
Eq.~\eqref{eq:H_app2} then becomes
\be 
\small
\begin{split}
H&= J_k \sum_k  \lf[\, a_{k,U}^{\dagger} b_{k,D} + a_{k,D}^{\dagger} b_{k,U} + {\rm H.c.} \rg]\\
&+ \rho \cos\theta
\sum_k  \lf[- a_{k,U}^{\dagger} a_{k,U} + a_{k,D}^{\dagger} a_{k,D} - b_{k,U}^{\dagger} b_{k,U}+  b_{k,D}^{\dagger} b_{k,D}   \rg] \\
&+  \rho \sin\theta
\sum_k  \lf[ i a_{k,U}^{\dagger} a_{k,D} + i  b_{k,U}^{\dagger} b_{k,D} + {\rm H.c.} \rg] \\
&+ \mu \sum_k  \lf[a_{n,U}^{\dagger} a_{k,D} - b_{k,U}^{\dagger} b_{k,D} + {\rm H.c.} \rg]  
\end{split} 
\label{eq:H_app2_th}
\ee
and consequently, in Eq.~\eqref{eq:H_k_app3} the vector fields become 
${\bf B}_a = ( \rho \sin\theta , - \mu, \rho \cos\theta )$ and ${\bf B}_b = (\rho \sin\theta, \mu , \rho\cos\theta )$
%${\bf B}_a = (\mu, \rho\sin\theta , \rho \cos\theta )$,   ${\bf B}_b = (-\mu,  \rho\sin\theta , \rho \cos\theta)$. 
The Hamiltonian Eq.~\eqref{eq:H_k_app3} becomes
\be
\begin{split}
H &= \sum_k \lf[- J_x\, \Phi_k ^\dagger (\tau_x\otimes\sigma_y) \Phi_k  -
J_y\, \Phi_k ^\dagger (\tau_y\otimes\sigma_y) \Phi_k  \rg] \\
&+\sum_k  \lf[ 
\rho \cos\theta \, \Phi_k ^\dagger(\tau_0\otimes\sigma_z) \Phi_k
+\rho \sin\theta \, \Phi_k ^\dagger(\tau_0\otimes\sigma_y) \Phi_k
\rg]  \\
&-\sum_{k} 
\mu\, \Phi_k ^\dagger (\tau_z\otimes\sigma_y) \Phi_k  
\end{split}
\label{eq:H_k_app_th2}
\ee
In terms of the Dirac matrices $\mathcal{L}$, the Hamiltonian $H_M$ of the Yang monopole in Eq.~\eqref{eq:H_Yang_v3} turns 
\be
\begin{split}
H_M 
%&=\sum_k \Phi_k ^\dagger  \lf[ 
% - J_y L_1  
%+ \rho_y  L_2 
%-\mu L_3   
%+ \rho_x  L_4 
% -J_x L_5
% \rg] \Phi_k  \\
 &=\sum_k \Phi_k ^\dagger  \lf[ 
 \mathcal{B}_\mu \mathcal{L}_\mu  
 \rg] \Phi_k 
\end{split}
\label{eq:H_k_app_th4}
\ee 
for $\mathcal{B}=( - J_y , \rho \sin \theta,-\mu,\rho\cos \theta, -J_x)$. 
%for $\mathcal{B}=( - J_y , \rho_y,-\mu,\rho_x,-J_x)$, $\rho_x = \rho \cos\theta$ and $\rho_y = \rho \sin\theta$. 

\end{document}